\documentclass[conference,a4paper]{IEEEtran}

\IEEEoverridecommandlockouts

\usepackage[cmex10]{amsmath}
\usepackage{multirow,tabularx}
\usepackage{array,graphicx}
\newcommand{\mbf}[1]{\mathbf{#1}}
\DeclareMathOperator{\perm}{perm}
\newtheorem{lemma}{Lemma}
\newtheorem{proposition}{Proposition}

\begin{document}

\title{Density Evolution for SUDOKU codes on the Erasure Channel}

\author{\IEEEauthorblockN{Caroline Atkins and Jossy Sayir}
\IEEEauthorblockA{University of Cambridge}
\thanks{Funded in part by the European Research Council under ERC grant
agreement 259663 and by the FP7 Network of Excellence NEWCOM\#
under grant agreement 318306.}}

\maketitle

\begin{abstract}
Codes based on SUDOKU puzzles are discussed, and belief propagation
decoding introduced for the erasure channel. Despite the non-linearity
of the code constraints, it is argued that density evolution can be
used to analyse code performance due to the invariance of the code under
alphabet permutation. The belief propagation decoder for erasure channels
operates by exchanging messages containing sets of possible values.
Accordingly, density evolution tracks the probability mass functions of
the set cardinalities. The equations governing the mapping of those
probability mass functions are derived and calculated for variable and
constraint nodes, and decoding thresholds are computed for long SUDOKU
codes with random interleavers.
\end{abstract}

\section{Introduction}

The analogy between SUDOKU puzzles and low-density parity-check (LDPC)
has been widely noted (\cite{farrell2009,moon2006,moon2009}).
Both can be represented by a factor graph, where
the constraints for LDPC codes are linear, i.e., $\sum c_ix_i = 0$ where
the coefficients and sum are defined over a finite field, while for
SUDOKU the constraints are non linear, requiring all variable in a
constraint to have different values within a finite alphabet.
This analogy can be pursued further by investigating the code
defined by the set of valid SUDOKU grids and its properties with
respect to communication over noisy channels. This investigation
brings up a number of interesting questions, some of which we 
have been able to provide answers for, while others remain 
unsolved:
\begin{itemize}
\item what is the rate of a code defined as the set of valid
(solved) SUDOKU grids? (unsolved for general dimensions)
\item how are SUDOKU encoded, e.g., how can we map information
sequences to the set of valid SUDOKU grids (some solutions
proposed, not covered in the present paper)
\item can SUDOKU codes be decoded via Belief Propagation (BP)
for general ``soft'' channels?
(solved in \cite{moon2006}, covered here)
\item error performance of SUDOKU codes (unsolved in general)
\item BP for the erasure channel? (solutions covered here)
\item performance of BP for the erasure channel (covered here).
\end{itemize}
In the present paper, we liberate SUDOKU puzzles from their strict
definition on a $q\times q$ grid with row, column and subgrid constraints,
and allow puzzles to grow to any length, with each variable involved
in a numbder $d_v$ of constraints, and each constraint involving
a number $d_c$ of variables. This is similar to regular long LDPC codes.
The aim of the paper is to provide analysis of the performance of
long SUDOKU codes using density evolution, a technique originally
developed to analyse LDPC codes. As we will see, this is slightly
more difficult that one could expect because SUDOKU codes are non-linear
and hence have no all-zero codeword, and performance analysis for
a given transmitted codeword would be insufficient anyhow. It is
therefore quite surprising that density evolution can be used
after all. 

In the next section, we will explain the decoder operations for
a SUDOKU code. In Section~\ref{sec:de}, we will develop
density evolution for SUDOKU codes, and finally provide results,
a conclusion and remaining open problems in the last section.

\section{Decoder}

Belief propagation for non-binary codes over an alphabet of size $q$
normally operates on a factor graph by exchanging messages in the
form of $q$-ary  a-posteriori probability distributions for the
$q$-ary code symbols. This is also true of SUDOKU codes, and the
node operations for general channels can easily be derived using
Bayesian inference. We present these mappings here
without further discussion as
\begin{equation}
q_{ij} = \alpha_i \prod_{k\neq i} p_{kj}.
\label{eq:var-op}
\end{equation}
for variable nodes, where $p_{kj}$ is the $j$-th entry of
the $k$-th probability-valued incoming message to the variable
node, while $\alpha_i$ is the a-posteriori probability that
the variable takes on the $i$-th value given the channel observation.
For constraint nodes of degree equal to the alphabet size $q$, the mapping is
\begin{equation}
q_{ij} = \xi_i\sum_{(j_1,\ldots,j_i=j,\ldots,j_q) \in S_q} \;\;\prod_{k\neq i} p_{kj_k}.
\label{eq:constraint-op}
\end{equation}
where $\xi_i$ is a normalisation constant and $S_q$ is the
symmetric group on $\{1,2,\ldots,q\}$. This can 
also be written as
\begin{equation}
q_{ij} = \frac{p_{ij}}{\perm(P)} \perm(P_{ij}).
\label{eq:constraint-op-perm}
\end{equation}
where $\perm(A)$ is the Cauchy permanent of a matrix $A$, $P$ is
the $q\times q$ matrix of incoming messages, and $A_{ij}$ for
any matrix $A$ denotes the matrix obtained by removing the $i$-th
row and $j$-th column from $A$.\footnote{ 
Note that, while the relation between SUDOKU constraint node operations and the
Cauchy permanent was never formally established in \cite{moon2009}. it is the
reason why Sinkhorn can be used to solve SUDOKU puzzles. Sinkhorn will converge
towards a doubly stochastic output matrix, and (\ref{eq:constraint-op-perm}) 
makes it evident that the Bayesian optimal output message matrix of a constraint node is
doubly stochastic.
}
These rules can readily be extended
to constraint nodes of any positive degree $d_c\leq q$, where all variables in 
a constraint must take on different values (they
cover all values in $\{1,\ldots,q\}$ only when $d_c=q$).

A $q$-ary erasure channel with input variable $X$ defined
over the alphabet $\mathcal{X}$ of cardinality $q$, and output alphabet $Y$
defined over $\mathcal{X}\cup\{\varepsilon\}$ where $\varepsilon$ is the
erasure symbol, has transition probabilities
\[
\begin{cases}
P_{Y|X}(y|x) = 1-\delta & \mbox{ for }y=x,\mbox{ and }\\
P_{Y|X}(y|x) = \delta & \mbox{ for }y=\varepsilon.
\end{cases}
\]
Note that the corresponding a-posteriori distributions are
$P_{X|Y}(x|\varepsilon) = 1/q$ for all $x$, and $P_{X|Y}(x|y)$ is $1$
if $x=y$ and $0$ otherwise. Hence,
for an erasure channel, messages will start off as
either atomic distributions assigning a probability of 1 to the 
observed symbol and 0 to all others, or uniform distributions
on the complete alphabet. Node operations will only ever yield
uniform distributions, where these would normally narrow down
to a subset of the alphabet as certain symbols get excluded,
until all messages converge to atomic distribution in the event
that decoding is successful. 
Therefore, the message-passing algorithm using distributions
can be replaced by a message-passing algorithm where messages
are alphabet subsets, where
a subset message is equivalent to a distribution message
with a uniform distribution over the subset.

Let $m_{ch}$ be an incoming subset message to a variable node from the
channel, where $m_{ch} = \{i\}$ if the channel output is symbol $i$,
and $m_{ch} = \{1,\ldots,q\}$ if the channel output is an erasure.
Let $m_{c\to v}(k)$ represents the incoming subset message to the same
variable node along its $k$-th edge, where $m_{c\to v}(k) \subseteq \{1,\ldots, q\}$.
Translating (\ref{eq:var-op}) from distributions to subset notation yields,
for the $k$-th outgoing message $m_{v\to c}(k)$ of the variable node
\[
m_{v\to c}(k) = m_{ch} \cap \bigcap_{k'\neq k} m_{c\to v}(k'),
\]
i.e., the outgoing message is an extrinsic intersection of all
incoming messages, since any zeros in an incoming distribution
would eliminate all non-zero probabilities for the same symbol
in other incoming distributions.

For constraint nodes, the relation between the distribution message operation
(\ref{eq:constraint-op}) and the subset operation is slightly more intricate.
The output $q_{ij}$ is zero if there exists a restriction of the matrix $P$
of incoming messages to the set of rows $\mathcal{I}'=\{i'_1,\ldots,i'_k\}$ not including $i$ whose
non-zero entries form a square matrix over the set of columns $\mathcal{J}'=\{j'_1,\ldots,j'_k\}$
including $j$. This square sub-matrix constitutes a bottleneck in the 
matrix $P$ so that every non-zero product of elements $\prod_k p_{kj_k}$
for a permutation $(j_1,\ldots,j_q)\in S_q$ must pass through
the bottleneck, i.e., $k\in\mathcal{I}'\implies j_k\in\mathcal{J}'$.
Therefore all of the terms in the expression (\ref{eq:constraint-op}) for $q_{ij}$ are zero when
$i\notin\mathcal{I}'$ and $j\in\mathcal{J}'$. 
Translated in terms of subset messages, square submatrices occur whenever the union of $k$ incoming
subset messages has cardinality $k$. 
Letting $m_{v\to c}(i) \subseteq \{1,\ldots,q\}$ be the incoming subset message to a constraint
node on its $i$-th edge, the resulting constraint node rule for generating the $j$-th
outgoing message $m_{c\to v}(j) \subseteq \{1,\ldots,q\}$ is
\[
m_{c\to v}(j) = \{1,\ldots,q\} - \bigcup_n A_n
\]
where $A_n$ is any set such that
\[
\exists \mathcal{J}\subset \{1,\ldots,q\} \mbox{ such that }
\begin{cases} k\notin\mathcal{J}, \\
A_n=\bigcup_{j\in\mathcal{J}} m_{v\to c}(j), \\
\mbox{ and } \#\mathcal{J} =  \#A_n,
\end{cases}
\]
where $\#S$ denotes the cardinality of the set $S$.

It is worth noting that the subset operations just stated are familiar
to passionate Sudoku solvers and are described in Sudoku solving guides
as the basic rules for solving most puzzles by logic excluding the fiendishly
difficult ones that require constraint combination tricks or guessing. The constraint
rule above with $\#A_n=1$ correspond to the simple rule of eliminating
any candidate value for a cell that is already used by another cell in 
the same constraint. The same rule for $\#A_n=2$ corresponds to
eliminating pairs of values that are shared among two other cells in the
same constraint, and so forth. Hence, we would claim that most
people who enjoy solving Sudoku puzzles are in fact running a
belief propagation algorithm for erasure channels in their brains.

\section{Density Evolution for SUDOKU codes}
\label{sec:de}

Density evolution (\cite{ru2001}) for LDPC codes relies on a number
of properties that generalise to the codes with SUDOKU-type constraints
considered here. In particular, concentration and convergence to the
cycle-free (tree-like) case apply to any code with local constraints that
can be represented as a factor graph, and this is the case for the SUDOKU-type
codes. However, one of the crucial properties that simplifies density evolution
for LDPC codes does not generalise to SUDOKU codes: the sufficiency of the 
analysis given the all-one codeword. Indeed, the all-one sequence is no
longer a codeword for the SUDOKU case. Indeed, even for optimal maximum likelihood
(ML) decoding, let alone sub-optimal iterative decoding, performance analysis given
a specific codeword does not suffice in general for non-linear codes, as the 
weight-distance equivalence property of linear codes does not extend to non-linear
codes. 

In principle, in order to make binding statements about the performance of
iterative decoding for SUDOKU codes, one would need to compute separate density
evolution recursions for every possible pattern of transmitted code symbols.
However, the following lemma listing the symmetries that all node operations
fulfills and allows us to overcome this hurdle:
\begin{lemma}
Consider a node of degree $d$ in a factor graph and consider an extrinsic
mapping from any $d-1$ of its input messages to the remaining output message, 
when decoding for the $q$-ary erasure channel. 
This mapping is invariant under the following:
\begin{itemize}
\item any of the $(d-1)!$ re-orderings of its input messages
\item any of the $q!$ permutations of the code alphabet and corresponding
  re-shuffling of the subset-valued intput and output messages
\end{itemize}
\label{lemma:invariance}
\end{lemma}
The lemma allows us to make the following simplifying assumptions
when computing the density evolution recursion:
\begin{itemize}
\item for a variable node, assume that the transmitted variable has value 1. Due to 
  the nature of the erasure decoder, this implies that the subset-valued messages
  from the channel and in- and outcoming along all edges to a variable node will
  contain at least the value 1. 
\item for a constraint node, assume that the transmitted variables corresponding
  to input edges 1 to $d-1$ have values 1 to $d-1$ respectively, and the transmitted variable
  corresponding to the output message has value $d$.
\end{itemize}
Despite the simplifications following from Lemma~\ref{lemma:invariance}, 
the message alphabet for density evolution is still rather large.
The following proposition that also follows from Lemma~\ref{lemma:invariance}
enables us to operate density evolution on a considerably reduced
message alphabet:
\begin{proposition}
For performance analysis, the probability distribution of the {\em cardinalities}
of subset-valued messages is a sufficient statistic for the probability
distribution of the messages themselves.
\end{proposition}
Hence, it will be sufficient to track the cardinality of messages, assuming
for variable nodes that each message contains at least the value 1, and for
constraint nodes that incoming messages contain the values 1 to $d-1$, respectively,
and the output message the value $d$. The cardinalities have value between 
1 and $q$ (value zero corresponding to the empty subset can never occur for
an erasure channel as the channel makes no mistake.)

The actual mappings of cardinalities are non-trivial and still subject
to a combinatorial explosion with growing $q$ and $d$. We will discuss
the computation of these mappings in the next two subsections, and then
present results for $q=d$ and $q$ between 3 and 6. Note that we currently have
the computing power to go up to $q=8$ and hope to find further
efficient implementation that would allow us to push the boundary to $q=9$,
but this is the limit beyond which the combinatorial explosion of terms
in the density evolution recursion would probably be beyond anyone's
computational abilities.

\subsection{Variable Node Equations}

Consider a variable node of degree $d_v$ operating over an
alphabet of size $q$. We now consider the extrinsic mapping
of messages from $d_v-1$ inputs to one output. We are
assuming incoming messages and hence their cardinalities to be
independent and have identical distributions of cardinality $P_{Vi}(k)$
for  $k=1,\ldots, q$. Note that $P_{Vi}(0)=0$ because any 
message always contains at least the corresponding transmitted symbol
for an erasure channel (the channel makes no ``errors''). 

Density evolution aims to express the probability distribution
of the cardinality of the output message $P_{Vo}(.)$ in function
of $P_{Vi}(.)$. In line with the assumptions above, we can 
assume without loss of generality that the true value of the
variable is 1 and hence all input messages and the output
messages must necessarily contain a 1. 

The density evolution analysis of a variable node is best
first illustrated with an example. Take $q=4$, $d_v=3$. 
In order to help avoid confusion, we will use bold numbers,
e.g., $\mbf{1}.\mbf{2},\ldots$, to denote set cardinalities and
normal numbers, e.g., $1,2,\ldots$ to denote possible values
of variables in a SUDOKU code.
There are 16 combinations of cardinalities for the two
input messages, i.e., 
$(\mbf{1},\mbf{1}), (\mbf{1},\mbf{2}),\ldots, (\mbf{2},\mbf{1}),\ldots$. 
Clearly we can restrict our attention to non-decreasing
combinations since the two input edges are essentially 
interchangeable. The number of non-decreasing combinations
for general $q$ and $d_v-1$, which we call $\mathcal{N}(q,d_v-1)$,
 can be defined recursively as
\[
\begin{cases}
\mathcal{N}(a,b) = \sum_{k=1}^a \mathcal{N}(k,b-1), \\
\mathcal{N}(a,1) = a.
\end{cases}
\]
In our example, this gives $\mathcal{N}(4,2) = 10$ as
can easily be verified by listing them: 
$(\mbf{1},\mbf{1})$, $(\mbf{1},\mbf{2})$, $(\mbf{1},\mbf{3})$, $(\mbf{1},\mbf{4})$,
$(\mbf{2},\mbf{2})$,$(\mbf{2},\mbf{3})$,$(\mbf{2},\mbf{4})$,$(\mbf{3},\mbf{3})$,
$(\mbf{3},\mbf{4})$, and $(\mbf{4},\mbf{4})$. Let us 
take the combination $(\mbf{2},\mbf{3})$ as an example. Since both
input messages must contain a 1, this implies a uniform
distribution over the sets $\{1,2\},\{1,3\}$, and $\{1,4\}$
for the message of cardinality 2, and a uniform distribution
over the messages $\{1,2,3\},\{1,2,4\}$ and $\{1,3,4\}$ for
the message of cadinality 3. There are 9 possible combinations
of those, 3 of which will yield the output message $\{1\}$ of
cardinality 1, and 6 of which will yield an output message
of cardinality 2, either $\{1,2\},\{1,3\}$ or $\{1,4\}$.
These probabilities can be added to obtain the overall 
distribution of output cardinalities, taking care to multiply
the resulting probabilities by the number of combinations, e.g.,
the combination of cardinalities $(\mbf{2},\mbf{3})$ counts double for
$(\mbf{2},\mbf{3})$ or $(\mbf{3},\mbf{2})$ while the combination of
cardinalities $(\mbf{2},\mbf{2})$
counts only for itself. This process is best illustrated with 
tables. Table~\ref{tbl:4-3-vnode} shows the probabilities of output
cardinalities given combinations of input cardinalities
and the corresponding multiplicity factor
\begin{table}[h!]
\centering
\begin{tabular}{|l|c||c|c|c|c|}
\hline
 & & \multicolumn{4}{c|}{output \#} \\
input \# & multipl. & $\mbf{1}$ & $\mbf{2}$ & $\mbf{3}$ & $\mbf{4}$ 
\\ \hline \hline
$(\mbf{1},\mbf{1})$ & 1 & 1 & 0 & 0 & 0 \\ \hline
$(\mbf{1},\mbf{2})$ & 2 & 1 & 0 & 0 & 0 \\ \hline
$(\mbf{1},\mbf{3})$ & 2 & 1 & 0 & 0 & 0 \\ \hline
$(\mbf{1},\mbf{4})$ & 2 & 1 & 0 & 0 & 0 \\ \hline
$(\mbf{2},\mbf{2})$ & 1 & 2/3 & 1/3 & 0 & 0 \\ \hline
$(\mbf{2},\mbf{3})$ & 2 & 1/3 & 2/3 & 0 & 0 \\ \hline
$(\mbf{2},\mbf{4})$ & 2 & 0 & 1 & 0 & 0 \\ \hline
$(\mbf{3},\mbf{3})$ & 1 & 0 & 2/3 & 1/3 & 0 \\ \hline
$(\mbf{3},\mbf{4})$ & 2 & 0 & 0 & 1 & 0 \\ \hline
$(\mbf{4},\mbf{4})$ & 1 & 0 & 0 & 0 & 1 \\ \hline
\end{tabular}
\caption{Output probabilities and multiplicities for $q=4$ and $d_v=3$}
\label{tbl:4-3-vnode}
\end{table}
The entries in the non-trivial columns in Table~\ref{tbl:4-3-vnode}
corresponding to input configurations $(\mbf{2},\mbf{2})$,
$(\mbf{2},\mbf{3})$ and $(\mbf{3},\mbf{3})$
can be visualised in the following tables listing all possible
pairs of inputs with these cardinalities and the corresponding
output cardinality,
\begin{center}
\begin{tabular}{|c||c|c|c|}
\hline
$(\mbf{2}, \mbf{2})$ & $\{1,2\}$ & $\{1,3\}$ & $\{1,4\}$ \\ \hline \hline
$\{1,2\}$ & $\mbf{2}$ & $\mbf{1}$ & $\mbf{1}$ \\ \hline
$\{1,3\}$ & $\mbf{1}$ & $\mbf{2}$ & $\mbf{1}$ \\ \hline
$\{1,4\}$ & $\mbf{1}$ & $\mbf{1}$ & $\mbf{2}$ \\ \hline 
\end{tabular}

\medskip
\begin{tabular}{|c||c|c|c|}
\hline
$(\mbf{2}, \mbf{3})$ & $\{1,2,3\}$ & $\{1,2,4\}$ & $\{1,3,4\}$ \\ \hline \hline
$\{1,2\}$ & $\mbf{2}$ & $\mbf{2}$ & $\mbf{1}$ \\ \hline
$\{1,3\}$ & $\mbf{2}$ & $\mbf{1}$ & $\mbf{2}$ \\ \hline
$\{1,4\}$ & $\mbf{1}$ & $\mbf{2}$ & $\mbf{2}$ \\ \hline 
\end{tabular}

\medskip
\begin{tabular}{|c||c|c|c|}
\hline
$(\mbf{3}, \mbf{3})$ & $\{1,2,3\}$ & $\{1,2,4\}$ & $\{1,3,4\}$ \\ \hline \hline
$\{1,2,3\}$ & $\mbf{3}$ & $\mbf{2}$ & $\mbf{2}$ \\ \hline
$\{1,2,4\}$ & $\mbf{2}$ & $\mbf{3}$ & $\mbf{2}$ \\ \hline
$\{1,3,4\}$ & $\mbf{2}$ & $\mbf{2}$ & $\mbf{3}$ \\ \hline 
\end{tabular}
\end{center}
Finally, the resulting distribution of output cardinalities can be
read out directly from Table~\ref{tbl:4-3-vnode} to yield
\begin{align*}
P_{vo}(\mbf{1}) = &\left(P_{vi}(\mbf{1})\right)^2 
+ 2P_{vi}(\mbf{1})\left(P_{vi}(\mbf{2})+P_{vi}(\mbf{3})+P_{vi}(\mbf{4})\right) \\
& +\frac{2}{3}\left(P_{vi}(\mbf{2})\right)^2 
 + \frac{2}{3}P_{vi}(\mbf{2})P_{vi}(\mbf{3}), \\
P_{vo}(\mbf{2}) = &\frac{1}{3}\left(P_{vi}(\mbf{2})\right)^2 
 + \frac{4}{3}P_{vi}(\mbf{2})P_{vi}(\mbf{3}) 
 + 2P_{vi}(\mbf{2})P_{vi}(\mbf{4}) \\
& + \frac{2}{3} \left(P_{vi}(\mbf{3})\right)^2, \\
P_{vo}(\mbf{3}) = & \frac{1}{3} \left(P_{vi}(\mbf{3})\right)^2 
+ 2P_{vi}(\mbf{3})P_{vi}(\mbf{4}), \\
P_{vo}(\mbf{4}) = & \left(P_{vi}(\mbf{4})\right)^2.
\end{align*}

The whole process can be summarized and generalized to any $q$ and
$d_v$ as follows
\[
P_{Vo}(k) =
\sum_{\mathbf{j}:j_2\leq j_3\leq \ldots j_{d_v}} 
\Gamma(\mathbf{j})
P_{Vo|Vi_1\ldots Vi_{d_v-1}}(k|j_2\ldots j_{d_v}) 
\]
where
\[
\Gamma(x_1,\ldots,x_n)=
\frac{n!}{\prod_i\#\{x_m=i\}!}
\]
where $\#$ denotes the cardinality of a set.
This can be further developed to give
\[
P_{Vo}(k) =
\sum_{\mathbf{j}:j_2\leq \ldots j_{d_v}} 
\Gamma(\mathbf{j})
\frac{\#\{m_{vo}:\#m_{vo}=k\}}{\#\{m_{vo}\}}
\prod_{m=2}^{d_v} P_{Vi}(j_m) 
\]
where the cardinalities of output message sets in the
fraction are sets of possible output messages $m_{vo}$
given all possible input messages of the cardinalities given by
$\mathbf{j}=(j_2,\ldots,j_{d_v})$.

\subsection{Constraint Node Equations}

Consider a constraint node of degree $d_c$ operating over an alphabet
of size $q$.  
For a classical SUDOKU puzzle, we have $d_c = q$, i.e., each
constraint ties $q$ variables to a permutations of the numbers
$1$ to $q$, but we can also consider the more general case
where $d_c\neq q$ and each constraint requires $d_c$ variables
to take on distinct values among the numbers 1 to $1$.
We consider the mapping of messages from $d_c-1$ inputs to one output.
We are assuming incoming messages and hence their cardinalities to be
independent and have identical distributions of cardinality
$P_{ci}(k)$ for $k=1,\ldots, q$. Note that $P_{ci}(0)=0$ because any
message always contains at least the corresponding transmitted symbol
for an erasure channel (the channel makes no ``errors'').

We will again aim to express the probability distribution of the
cardinality of the output message $P_{co}(.)$ in function of
$P_{ci}(.)$.  In line with the assumptions above, we can assume
without loss of generality that the oputput message goes to a variable
with true value 1.  The output message must therefore contain a 1.
The input messages each come from variables with different true
values.  The messages must each contain the value held by their source
variable.  Without loss of generality we can assume these values to be
$2,\ldots, d_c$.

As for variable nodes in the previous section, the constraint node
can again best be demonstrated with an example.  Take
the case where $q=4$ and $d_c = 4$.  The same conventions are used as
with the variable node.  There are 64 combinations of input
cardinalities.  Again these can be grouped in non-decreasing
combinations.  There are 20 such combinations.

Let us take the combination $(\mbf{1}, \mbf{2},\mbf{3})$ as an
example.  The three messages must contain the values of their source node.
This implies that the message of cardinality $\mbf{1}$ is $\{2\}$.
There is a
uniform distribution over messages $\{1,3\}, \{2,3\}$, and $\{3,4\}$
for the message of cardinality $\mbf{2}$, and a uniform distribution over
messages $\{1,2,4\}, \{1,3,4\}$, and $\{2,3,4\}$ for the message of
cardinality $\mbf{3}$. 
There are 9 possible combinations of these.  Two will
yield output message $\{1\}$, these are $\{2\}$,$\{2,3\}$,$\{2,3,4\}$, and
$\{2\}$,$\{3,4\}$,$\{2,3,4\}$. Two will yield the output message
$\{1,4\}$ of cardinality $\mbf{2}$, these are $\{2\}$,$\{2,3\}$,
$\{1,2,4\}$ and
$\{2\}$,$\{2,3\}$,$\{1,3,4\}$.  The remaining five combinations yield
output message $\{1,3,4\}$ of cardinality 3.

\begin{table}[h!]
\centering
\begin{tabular}{|l|c||c|c|c|c|}
\hline
 & & \multicolumn{4}{c|}{output \#} \\
input \# & multipl. & $\mbf{1}$ & $\mbf{2}$ & $\mbf{3}$ & $\mbf{4}$ 
\\ \hline \hline
$(\mbf{1},\mbf{1},\mbf{1})$ & 1 &       1 &      0 &      0 &       0 \\ \hline
$(\mbf{1},\mbf{1},\mbf{2})$ & 3 &       2/3 &    1/3 &    0 &       0 \\ \hline
$(\mbf{1},\mbf{1},\mbf{3})$ & 3 &       1/3 &    2/3 &    0 &       0 \\ \hline
$(\mbf{1},\mbf{1},\mbf{4})$ & 3 &       0 &      1 &      0 &       0 \\ \hline
$(\mbf{1},\mbf{2},\mbf{2})$ & 3 &       4/9 &    2/9 &    1/3 &     0 \\ \hline
$(\mbf{1},\mbf{2},\mbf{3})$ & 6 &       2/9 &    2/9 &    5/9 &     0 \\ \hline
$(\mbf{1},\mbf{2},\mbf{4})$ & 6 &       0 &      1/3 &    2/3 &     0 \\ \hline
$(\mbf{1},\mbf{3},\mbf{3})$ & 3 &       1/9 &    0 &      8/9 &     0 \\ \hline
$(\mbf{1},\mbf{3},\mbf{4})$ & 6 &       0 &      0 &      1 &       0 \\ \hline
$(\mbf{1},\mbf{4},\mbf{4})$ & 3 &       0 &      0 &      1 &       0 \\ \hline
$(\mbf{2},\mbf{2},\mbf{2})$ & 1 &       8/27 &   1/9 &    0 &       16/27 \\ \hline
$(\mbf{2},\mbf{2},\mbf{3})$ & 3 &       4/27 &   2/27 &   0 &       21/27 \\ \hline
$(\mbf{2},\mbf{2},\mbf{4})$ & 3 &       0 &      1/9 &    0 &       8/9 \\ \hline
$(\mbf{2},\mbf{3},\mbf{3})$ & 3 &       2/27 &   0 &      0 &       25/27 \\ \hline
$(\mbf{2},\mbf{3},\mbf{4})$ & 6 &       0 &      0 &      0 &       1 \\ \hline
$(\mbf{2},\mbf{4},\mbf{4})$ & 3 &       0 &      0 &      0 &       1 \\ \hline
$(\mbf{3},\mbf{3},\mbf{3})$ & 1 &       1/27 &   0 &      0 &       26/27 \\ \hline
$(\mbf{3},\mbf{3},\mbf{4})$ & 3 &       0 &      0 &      0 &       1 \\ \hline
$(\mbf{3},\mbf{4},\mbf{4})$ & 3 &       0 &      0 &      0 &       1 \\ \hline
$(\mbf{4},\mbf{4},\mbf{4})$ & 1 &       0 &      0 &      0 &       1 \\ \hline
\end{tabular}
\caption{Output probabilities and multiplicities for $q=4$ and $d_c=4$}
\label{tbl:4-4-cnode}
\end{table}

Finally, the resulting output cardinalities can be expressed as
\begin{align*}
P_{co}(\mbf{1}) = &\left(P_{ci}(\mbf{1})\right)^2 
+2P_{ci}(\mbf{1})^2P_{ci}(\mbf{2})
+2P_{ci}(\mbf{1})^2P_{ci}(\mbf{3}) \\
&
+\frac{4}{3}P_{ci}(\mbf{1})P_{ci}(\mbf{2})^2
+\frac{4}{3}P_{ci}(\mbf{1})P_{ci}(\mbf{2})P_{ci}(\mbf{3}) \\
&
+\frac{1}{3}P_{ci}(\mbf{1})P_{ci}(\mbf{3})^2
+\frac{8}{27}P_{ci}(\mbf{2})^3
+\frac{4}{9}P_{ci}(\mbf{2})^2P_{ci}(\mbf{3}) \\
&
+\frac{2}{9}P_{ci}(\mbf{2})P_{ci}(\mbf{3})^2
+\frac{1}{27}P_{ci}(\mbf{3})^3 
\end{align*}
\begin{align*}
P_{co}(\mbf{2}) = & P_{ci}(\mbf{1})^2P_{ci}(\mbf{2})
+2P_{ci}(\mbf{1})^2P_{ci}(\mbf{3})
+3P_{ci}(\mbf{1})^2P_{ci}(\mbf{4})\\
&
+\frac{2}{3}P_{ci}(\mbf{1})P_{ci}(\mbf{2})^2
+\frac{4}{3}P_{ci}(\mbf{1})P_{ci}(\mbf{2})P_{ci}(\mbf{3})\\
&
+2P_{ci}(\mbf{1})P_{ci}(\mbf{2})P_{ci}(\mbf{4})
+\frac{1}{9}P_{ci}(\mbf{2})^3 \\
&
+\frac{2}{9}P_{ci}(\mbf{2})^2P_{ci}(\mbf{3}) 
+\frac{1}{3}P_{ci}(\mbf{2})^2P_{ci}(\mbf{4}) \\
P_{co}(\mbf{3}) = & P_{ci}(\mbf{1})P_{ci}(\mbf{2})^2
+\frac{10}{3}P_{ci}(\mbf{1})P_{ci}(\mbf{2})P_{ci}(\mbf{}) \\
& +4P_{ci}(\mbf{1})P_{ci}(\mbf{2})P_{ci}(\mbf{4})
+\frac{8}{3}P_{ci}(\mbf{1})P_{ci}(\mbf{3})^2 \\
&+6P_{ci}(\mbf{1})P_{ci}(\mbf{3})P_{ci}(\mbf{4})
+3P_{ci}(\mbf{1})P_{ci}(\mbf{4})^2 \\
P_{co}(\mbf{4}) = & \frac{16}{27}P_{ci}(\mbf{2})^3
+\frac{7}{3}P_{ci}(\mbf{2})^2P_{ci}(\mbf{3})
+\frac{8}{3}P_{ci}(\mbf{2})^2P_{ci}(\mbf{4})\\
&
+\frac{25}{9}P_{ci}(\mbf{2})P_{ci}(\mbf{3})^2
+6P_{ci}(\mbf{2})P_{ci}(\mbf{3})P_{ci}(\mbf{4}) \\
&
+3P_{ci}(\mbf{2})P_{ci}(\mbf{4})^2
+\frac{26}{27}P_{ci}(\mbf{3})^3 \\
&
+3P_{ci}(\mbf{3})^2P_{ci}(\mbf{4})
+3P_{ci}(\mbf{3})P_{ci}(\mbf{4})^2
P_{ci}(\mbf{4})^3.
\end{align*}

\subsection{Results and Discussion}

The density evolution recursions outlined above can be used to compute thresholds
for long SUDOKU-type codes in a similar fashion as is done for LDPC codes. Convergence
to cardinality 1 message is the equivalent to error-free decoding, and the threshold
is the limit between error-free decoding and values of the erasure probability for
which the decoding error after any number of iterations remains positively lower bounded.

However, one major difficulty for SUDOKU codes is that the code rate is unknown
and determining it remains an open problem. We do have a conjecture but are
unable to give it a full justification at this point: we obtain a rate estimate by counting
the number of possible values that a set of variables can take on when those variables
are arranged in the tree resulting from considering the decoding neighbourhood of one
node in a finite number of iterations within the bi-partite factor graph corresponding
to a SUDOKU code. For a regular $(d_v,d_c)$ SUDOKU code of alphabet size $q$,
this results after $k$ iterations, in a rate
\[
R_k = \frac{\log_q\left(d_c!((d_c-1)!)^{k(d_v-1)}\right)}{d_c+k(d_c-1)(d_v-1)}
\]
which, as $k$ grows large, tends towards
\begin{equation}
R = \frac{\log_q((d_c-1)!)}{d_c-1}.
\label{eq:rate}
\end{equation}
It is somewhat surprising that our final conjectured rate for long codes
does not depend on the variable
degree of the nodes but only on the constraint node degree $d_c$ and on the
alphabet size $q$. This may indicate a weakness of our construction but
more analysis is needed to fully understand whether the number we calculate
here is close to the actual rate or merely an upper bound.

In Table~\ref{table:thresholds}, we list the thresholds (in terms or erasure probability)
calculated through densitye
evolution and the conjectured estimated rates resulting from (\ref{eq:rate}).
\begin{table}
\centering
\begin{tabular}{|c|c|c||c|c|}
\hline
$q$ & $d_v$ & $d_c$ & $\theta_{de}$ & $R$ \\ \hline \hline
3 & 3 & 3 & 0.98426 & 0.3155\\ \hline
4 & 3 & 4 & 0.94142 & 0.4308\\ \hline
5 & 3 & 5 & 0.89843 & 0.4937\\ \hline
6 & 3 & 6 & 0.86026 & 0.5344\\ \hline
\end{tabular}
\caption{Threshold vs. conjectured rate for long SUDOKU codes}
\label{table:thresholds}
\end{table}
The table indicates that there is a wide gap between threshold and rate at this point,
where the gap appears to become thinner as the alphabet size grows. However, it should
be noted that
the rate estimate is only a conjecture at this point assuming that the whole
inifite length codeword can be represented as a tree-like decoding neighbourhood.
More theory is needed to test this assumption. Intuitively, we expect the rates currently
provided to be upper bounds on the true rate.

\section{Conclusion}

We have presented the essential components of an erasure iterative decoder for locally
decodable codes fulfilling SUDOKU-type constraints. We have shown that density evolution
for this type of channel can be simplified to tracking the probability distribution
of the message cardinalities, effectively a $q$-ary probability vector instead of the
$2^q$ probability vector that would be required to operate density evolution on 
the full message alphabet. We have shown some preliminary numerical results, listing
thresholds that emerge from the density evolution recursion, alongside
a conjectured estimate for the rate of long SUDOKU-type codes.

Furter work will require a firmer grasp on the code rate and comparison of the
thresholds for other code dimensions. Also, a simulation of code performance for
various block lengths would be of interest but full implementation of an encoder
and decoder for SUDOKU-type codes still requires solutions of some unsolved technical
details, as mentioned in the introduction. 



%



\bibliographystyle{IEEEtran}

\end{document}